\documentclass[conference]{IEEEtran}
\usepackage{cite}
\usepackage{amsmath, amssymb}
\usepackage{physics}
\usepackage{qcircuit}

\usepackage{graphicx} \usepackage{subcaption}

\usepackage{xcolor}

\usepackage[all]{nowidow}

\definecolor{W_state}{RGB}{176, 224, 230} 
\definecolor{GHZ_state}{RGB}{70, 130, 180} 
\definecolor{HR}{RGB}{255, 105, 180} 
\definecolor{BC}{RGB}{255, 127, 80} 
\definecolor{RC}{RGB}{100, 149, 237} 
\definecolor{UCNOT}{RGB}{0, 0, 205} 
\definecolor{iQFT}{RGB}{138, 43, 226}

\title{Randomized and Diverse Input State Generation for Quantum Program Testing}

\author{
\IEEEauthorblockN{Maryse Ernzer}
\IEEEauthorblockA{\textit{University of Luxembourg}\\
Luxembourg, Luxembourg \\
maryse.ernzer@uni.lu}
\and
\IEEEauthorblockN{Seung Yeob Shin}
\IEEEauthorblockA{\textit{University of Luxembourg}\\
Luxembourg, Luxembourg \\
seungyeob.shin@uni.lu}
\and
\IEEEauthorblockN{Fabrizio Pastore}
\IEEEauthorblockA{\textit{University of Luxembourg}\\
Luxembourg, Luxembourg \\
fabrizio.pastore@uni.lu}
\and
\IEEEauthorblockN{Domenico Bianculli}
\IEEEauthorblockA{\textit{University of Luxembourg}\\
Luxembourg, Luxembourg \\
domenico.bianculli@uni.lu}
}

\date{February 2026}

\begin{document}

\maketitle

\begin{abstract}
With the accelerating development of quantum technologies and their growing computational potential, quantum systems are being adapted for simulations and other critical tasks across diverse domains, making the reliability of the corresponding quantum software an essential concern.  Although recent efforts have started to incorporate quantum‑specific properties such as magnitude, phase, and entanglement  under the form of input‑coverage criteria into software testing, the unique structure of the quantum state space demands for more comprehensive testing. In particular, the notion of complete state-space exploration has so far received little attention.

To address this gap, we propose a framework for evaluating test circuit generators with respect to their coverage of the quantum state space. 
Our contribution is threefold: we develop a set of diversity scores that capture both local and global indicators of the extent to which the state space is explored; we propose a test circuit generator that produces test input states via a Brick-Circuit (BC) construction designed to approximate ideal random states using hardware-compatible gates;
we compare the proposed construction with existing generators based on their ability to generate uniformly distributed random test input states.

Our extended diversity scores quantify the local correlations and global spread of magnitude, phase and entanglement.
Using these scores, we evaluate the expressibility, defined as the capability to span the quantum state space uniformly, and entangling capabilities of existing generators relative to the BC generator.
Our results show that the hardware‑compatible BC generator achieves higher expressibility and entanglement performance at shallower depths than existing circuit generators. 

\end{abstract}

\section{Introduction}
With the continued progress of quantum technologies, software development evolves in parallel with its hardware counterpart. Given the novelty of the field, quantum software is particularly susceptible to programming mistakes, underscoring the need for systematic approaches to reliability. 
Additionally, the inherent characteristics of quantum systems, such as probabilistic outcomes, sensitivity to measurement, and the exponentially large state space, impose challenges on the testing methods that go beyond those of classical systems and demand for adapted testing techniques~\cite{pezze20252030}. 

Recent approaches aim to address these challenges by exploiting quantum-specific properties such as superpositions, entanglement and reversibility~\cite{honarvar2020property, abreu2022metamorphic, pontolillo2025qucheck}.
These works indicate that incorporation of quantum-specific properties improves fault detection capabilities, especially by uncovering quantum-specific faults~\cite{miranskyy2025feasibility, huang2019statistical, li2020projection, oldfield2025faster}.

To further enhance fault detection and overall testing effectiveness in a systematic manner, test case diversity has been investigated. 
Early work introduced coverage criteria and test‑input optimization methods, along with systematic frameworks, search‑based techniques incorporating quantum features and fuzzing‑based state‑space exploration for fault detection~\cite{ali2021assessing, wang2021quito, wang2021search, wang2021fuzz,long2024testing}. These studies also highlight the value of quantum‑aware input coverage, and demonstrate the effectiveness of including state superpositions and entanglement to ensure test thoroughness.
More recent approaches explicitly aim to quantify and increase the diversity of generated test inputs, addressing the challenge of achieving effective coverage of the quantum input state space and of generating test input states that exhibit genuinely quantum properties, which typically require nontrivial quantum circuit constructions~\cite{ye2023qura}.
In the latter, the local state properties (e.g., magnitude and phase) and non-local correlation properties (e.g., entanglement), are directly considered and formulated as diversity scores quantifying the property distributions across bipartite qubit sectors. These scores are used to measure the input coverage achieved by certain test circuit generators.
The test generators follow different patterns, including simple concatenation of random gates, a set of random rotation combined with entangling gates and a structured iQFT implementation.

While the methods described above address some manifestations of quantum states such as phase, superposition, and entanglement, a key aspect of quantum mechanics---namely the complex structure of the underlying state space---has so far been mostly overlooked in quantum software testing (QST).
Indeed, for multi-qubit systems, the dimensionality of the state space has to be considered in the diversity assessment, since generating test input states with a wide distribution of magnitude and phase does not prevent the state from being confined to a small subset of the entire (Hilbert) state volume.
To assess expressibility, defined by the ability to span the entire state space and generate entanglement, more informative indicators than magnitude and phase have to be considered.
For example, recent works have established a direct link between maximizing state‑space exploration and the uniformity of the generated states under the form of Haar‑distributed ideal random states~\cite{sim2019, nakaji2021expressibility, azado2025expressibility} that can be approximated via circuits that implement unitary-2 designs~\cite{emerson2003, harrow2009, dankert2009exact, brandao2016}.

We propose to advance existing QST approaches along these lines,
by integrating state-of-the-art techniques developed within the quantum information processing community~\cite{sim2019, nakaji2021expressibility}. Specifically, to ensure that test input states span a substantial part of the state-space volume, we propose to extend diversity scores such that they capture not only local features but also multi-qubit correlations, in order to establish a comprehensive expressibility profile of test circuit generators, which produce test input states. 
These scores include a general estimation of state participation and distinguish between structured subsystems and global properties, addressing the need for more fine‑grained, informative coverage measures in QST.
We verify their capability in capturing the targeted properties by applying them to well-defined reference quantum states.

Moreover, to guarantee high coverage through an unbiased exploration of the quantum state space, we implement an efficient test circuit generator based on the Brick-Circuit (BC) construction~\cite{emerson2003,harrow2009,brandao2016}, a quantum circuit pattern designed to approximate the Haar-random distribution using only nearest-neighbor 2-qubit entangling gates, ensuring feasibility on the current quantum processors. 
We compare the degree of randomness of the circuit matrices generated via the hardware-compatible BC construction to three different construction methods illustrated in the state of the art~\cite{ye2023qura}, namely the Random, UCNOT and iQFT designs.
We evaluate the generated randomness through the statistical proximity of generated unitary ensembles to ideal Haar-distributed samples. 
Our results show that our BC generator saturates the randomness threshold with fewer gates than previously used test circuit generators. We also observe that circuits constructed by combining only a few randomly drawn gates exhibit insufficient randomness.

Finally, we compare the proposed BC construction with existing test circuit generators, by analyzing the diversity of the test input states produced by their respective circuit designs, as quantified by our set of diversity scores, therby addressing current limitations in evaluating the coverage and expressibility of quantum circuit generators.
Therefore, we apply the extended diversity scores to quantify the performance of the circuit generators in expressibility and entangling capabilities. To this end, we consider qubit-sector occupations, local phase and amplitude correlations, effective state participation, as well as single-qubit and delocalized entanglement.
We find that the BC construction consistently produces test input states that reach the highest diversity scores, closest to the expected values of ideal random states.

In summary, the main contributions of this paper consist of the following.
\begin{itemize}
\item  We extend diversity scores to capture essential quantum state features. 
\item We investigate the BC construction, known for its efficient random state generation, as a test circuit generator for diverse test input states.
\item We evaluate different test circuit generators based on their expressibility and entangling capabilities.
\end{itemize}
Our results pave the way towards effective and comprehensive test circuit generation for QST. In particular, they combine a circuit generator that produces test input states which efficiently approximate Haar-random states with diversity scores that capture expressibility and entangling capabilities, enabling a more complete exploration of the admissible quantum state space during testing.

The rest of the paper is organized as follows: Section~\ref{sec:bg} provides some background information to understand the rest of the paper; section~\ref{sec:ap} presents the extended the diversity scores and the proposed test circuit generator.
In Section~\ref{sec:eval}, we evaluate and compare our developed diversity scores and test circuit generator with existing techniques. 
In Section~\ref{sec:rel} we review related work. Section~\ref{sec:con} draws conclusions and outlines future work. 
 \section{Background}
\label{sec:bg}
\subsection{The Quantum State}
\label{subsec:qs}

The general state of a $n$-qubit system belongs to the complex vector space $\mathbb{C}^d$ of dimension $d= 2^n$ growing exponentially in the qubit number $n$. 
Let $ \{\ket{j}\}_{j=0}^{n-1} = \left\{\ket{j_0 j_1 ...\,j_{n-1}} \lvert j_k \in \{0,1\}\right\}$ be the orthonormal computational basis where each state is identified with its binary expansion $j = \sum_{k=0}^{n-1} j_k 2^{n-1-k}$;
an arbitrary pure state can be described in terms of its \textit{state vector} $\ket{\psi} = \sum_{j = 0 }^{2^n -1} \alpha_j \ket{j} $ with the \textit{complex amplitudes} $\alpha_j =  r_j  \textrm{e}^{\textrm{i} \phi_j}$ associated with the state $\ket{j}$ which can be decomposed in terms of \textit{magnitude} $r_j = \lvert  \alpha_j \rvert$ and \textit{phase} $\phi_j \in [0,2\pi)$. The state is necessarily normalized by $ \sum_{j=0}^{2^n -1} \lvert \alpha_j\rvert^2 = 1 $, where the terms $\lvert \alpha\rvert_j^2$ express the probability of observing the basis state $\ket{j}$ upon measurement in the computational basis.

As a consequence of maintaining multi-qubit phase correlations within the complex state structure of quantum systems, non-classical correlation properties such as \textit{quantum entanglement} emerge. These result in a non-separability of multi-qubit states, i.e., the combined state of subsystems A and B cannot be written as the separable product of its constituents $\ket{\psi}_{AB} = \ket{\psi_A} \otimes \ket{\psi_B}$.

Illustrative examples representing two classes of strongly entangled states are given by the Greenberg-Horne-Zeilinger (GHZ) state incorporating cat-state entanglement and the W state  implementing  delocalized single‑excitation entanglement~\cite{briegel2001persistent, dur2000three}, defined as follows:
\begin{align}
    \psi_\mathrm{GHZ} &= \frac{1}{\sqrt{2}} ( \ket{0}^{\otimes n} + \ket{1}^{\otimes n}, \label{eq:ghz} \\
    \psi_\mathrm{W} &= \frac{1}{\sqrt{n}} \sum_{k=0}^{n-1} \ket{0 ... 0 \,1_k\, 0 ... 0}. \label{eq:w}
\end{align}

\subsection{Entanglement measures}
\label{sec:es}
Since quantum entanglement is a primary resource for quantum computation, various measures have been developed to quantify the entanglement strength. Such measures employ different approaches depending on the aspects they aim to capture and the system configuration; they have to satisfy monotonicity under Local Operations and Classical Communication (LOCC) and return $0$ if and only if the considered states are separable, and $1$ for maximally-entangled states.

We consider two measures that evaluate the strength of the entanglement via the presence of entropy in the subsystems. 
The \textit{von Neumann entropy} captures the average strength of the entanglement in the system in terms of 
the density matrix $\rho = \ket{\psi}\bra{\psi}$ of state $\ket{\psi}$, and is defined (\cite{bennett1996mixed,nielsen2010quantum}) as 
\begin{align}
S = - \mathrm{Tr} ( \rho \log \rho). \label{eq:entropy} 
\end{align}

An alternative entanglement measure consists in the \textit{R\'enyi-2 entropy}, which quantifies the entanglement between a subsystem A and its complement. It is defined (\cite{calabrese2004entanglement,nielsen2010quantum}) as 
\begin{align}
S_2(A) = - \textrm{log Tr} (\rho_A^2) \label{eq:r2e},
\end{align}
where $\rho_A$ is the reduced density matrix of the considered subsystem A. Unlike the von Neumann entropy, this measure does not require diagonalization and the knowledge of the full eigenvalue spectrum and is based solely on easily accessible measured quantities. As a result, it is particularly attractive for its numerical and experimental practicality~\cite{hastings2010measuring,islam2015measuring, zhao2022measuring}. Through its sensitivity to the chosen partition, the R\'enyi-2 entropy provides detailed insights into the extent and delocalization of entanglement in multipartite subsystems; moreover, since it does not rely on averaging, it is especially sensitive to extreme features in the entanglement spectrum.

Considering the global spread of entanglement across all qubits, a widely used measure consists of the \textit{Meyer-Wallach (MW) measure}~\cite{meyer2002global, brennen2003observable}. It returns the average value of the entanglement strength between each qubit and the rest of the system; it is defined as 
\begin{align}
    Q(\ket{\psi}) = 2 [ 1- \frac{1}{n} \sum_{k = 1}^n \mathrm{Tr} (\rho_k^2) ], \label{eq:qmw}
\end{align} 
for a $n$-qubit system and $\rho_k$ the reduced density matrix of the $k$th qubit. It is especially interesting for its computational efficiency, scalability and its ability to indicate the onset and local saturation of entanglement \cite{de2006multipartite}.
The average value of the MW  measure has been used to estimate the entangling capability of quantum circuits~\cite{sim2019}.

\subsection{Unitary Description}
The general closed evolution of quantum states is reversible and can be expressed by a \textit{unitary matrix} $U$, such that the resulting state can be written as $\ket{\psi }= U \ket{\psi_\mathrm{in} }$ in terms of the initial state $\ket{\psi_\mathrm{in}}$. Within the scope of quantum computing, this means that the combined effect of all gates acting on a target state can be summarized via the matrix product of the individual unitary representation acting on an initial state, which in the default case is given by $\ket{\psi_\mathrm{in}} = \ket{0}^{\otimes n}$. 

The \textit{unitary circuit} is the matrix obtained by multiplication of the applied gate sequence. Uniform sampling of the quantum state space therefore corresponds to sampling uniformly from the space of unitary matrices, i.e., drawing a random matrix.
A uniform distribution over all unitaries must be invariant under all unitary transformations. This property uniquely characterizes the \textit{Haar measure} $\mu$, defined by its left- and right-invariance~\cite{zyczkowski94}. Any matrix $M \in U(N)$ with the measurable subset $S \subseteq U(N)$ must meet $ \mu (SM) = \mu(MS) = \mu(M)$ with normalization $\int_U \mathrm{d} \mu(M) =1$.
Hence, the Haar measure assigns equal weight to all regions, which can be used to derive randomized states as explained in the following paragraph. 
Furthermore, the significant properties of Haar-random states that we consider, such as amplitude and phase correlations as well as entropy, are defined by at most second order statistical moments, and thus are captured by a unitary-2 design approximation~\cite{emerson2003, dankert2009exact, brandao2016}. 

\subsection{Randomized states}
For an n-qubit system, we can use  Haar matrix $U_\mathrm{Haar} \sim \mathrm{Haar}(U(2^n))$ to define differently \textit{structured random states}, such as a generic Haar-random state or randomized maximally-entangled bipartite state  
\begin{align}
    \ket{\psi_\mathrm{hr}} &= U_\mathrm{Haar} \ket{0}^{\otimes n}, \label{eq:haar}\\
    \ket{\psi_\mathrm{me}} &= (I_A \otimes U_B) \frac{1}{\sqrt{2^{n_A}}} \sum_{a = 0}^{2^{n_{A}}-1} \ket{a}_A \otimes \ket{a}_B, \label{eq:me}
\end{align}
with the identity matrix $I_A$ of the subsystem A of size $n_A$, and a Haar matrix $U_B \sim \mathrm{Haar} (U(2^{n_B}))$ applied to subsystem B of size $n_B= n-n_A$.
Further interesting classes of random states consist of uniform flat-amplitude states and random product states defined as
\begin{align}
\ket{\psi_\mathrm{ua}} &= \frac{1}{\sqrt{2^n}} \sum_{j = 0 }^{2^n-1} \mathrm{e}^{i \phi_j}\ket{j}, \label{eq:ua}\\
\ket{\psi_\mathrm{p}} &= \bigotimes_{k=1}^{n} (\cos{\frac{\theta_k}{2}} \ket{0} + \mathrm{e}^{i\phi_k} \sin{\frac{\theta_k}{2}}\ket{1}), \label{eq:p}
\end{align}
with random angles $\theta_k \in [0,\pi ],\phi_k \in [ 0,2\pi)$.

\subsection{Expressibility} 
\label{sec:express}
\textit{Expressibility} of quantum circuits refers to their ability to represent any possible state of the underlying complex Hilbert space.  Since this corresponds exactly to the capability of generating a uniform distribution of states, expressibility is directly related to the generation of Haar-random states.
Together with the entangling capability, expressibility has been identified as one of the two key circuit descriptors that provide a quantitative characterization of parametrized quantum circuits~\cite{sim2019}.
In the context of test-case generation, expressibility therefore captures the diversity of test input states with respect to the entire accessible state space, and has been directly linked to proximity to Haar-random states and the presence of entanglement \cite{sim2019, liu2022, mele2024}.

 \section{Diversity scores and test generation}
\label{sec:ap}
This section presents the six diversity scores that we are going to use for characterizing test input states with respect to their  magnitude and phase. 
Building on measures developed in statistical mechanics and chaos theory, we transfer and adapt a set of diversity scores to quantify the expressibility of test input states~\cite{zyczkowski94, fisher1995statistical,  mardia2009directional,ye2023qura, liu2025quantum, claeys2025fock}. 
Furthermore, we describe the test circuit generator that we adapted from studies in theoretical and experimental quantum physics~\cite{emerson2003, harrow2009, brandao2016, haferkamp2022random, schuster2025random}, to achieve our goal of efficiently producing test input states. 
\subsection{Diversity Scores}
To quantify the diversity of a state generated by a test circuit, we characterize the exploration of the quantum state space by a set of scores addressing the magnitude and phase of the individual state, analogous and complementary to the established entanglement measures described above. 

We start by considering the magnitude and phase scores originally introduced in QuraTest~\cite{ye2023qura}, which capture how evenly the magnitude and phase contributions are distributed 
between the ensembles of states forming two sectors defined by a given qubit being in state $0$ or $1$ with all other qubit entries kept fixed. Although this provides insight into the balance of magnitude and phase via one-to-one correlations between the two sectors, it neglects the more complex aspects of all-to-all correlations within the amplitude and phase distribution. 

To establish a comprehensive set of scores that describe the expressibility of a given state, we reformulate and formalize the original QuraTest scores in a unified framework, which we further complement with correlation and global spread scores.
We express the scores via the amplitudes $\alpha_j$ of the state vector $\ket{\psi}$ resulting from applying the circuit unitary $U$ to the default initial state $\ket{0}^{\otimes n}$. 
For each qubit $k$, we define a binary partition of the state space into two qubit sectors corresponding to the ensembles of computational basis states for which the $k$-th qubit is fixed to $0$ or $1$, respectively. 
Following the convention of notation introduced in Sec.~\ref{subsec:qs}, we define the sectors $S_0^k = \text{span}\{\ket{j} : j_k = 0 \} $ and $S_1^k =   \text{span} \{\ket{j}:  j_k = 1 \}$, where  ``span'' indicates the linear span operator. The vectors in opposite sectors are separated by the index value $p_k = 2^{n-1-k} $.
Furthermore, for each qubit k, we define the support for pairs of states with non-zero amplitude mass $\mathcal{I}_{\mathrm{p}}^k =  \{j : j_k =0, \; \lvert \alpha_j \rvert \neq 0, \;\lvert \alpha_{j+p_k} \rvert \neq 0  \} $  resulting in the number of pairs of contributing basis states $N_{\mathrm{p}}^k = \lvert \mathcal{I}_\mathrm{p}^k\rvert$, while the dimensions of the general support $\mathcal{I}_{}^k =  \{j: \lvert \alpha_j \rvert \neq 0\}$ is $N^k = \lvert \mathcal{I}_{}^k \rvert$, and of the support for the states either being 0 or 1, $\mathcal{I}_{0}^k =  \{j: j_k = 0,  \lvert \alpha_j \rvert \neq 0\}$ and $\mathcal{I}_{1}^k =  \{j: j_k = 1,  \lvert \alpha_j \rvert \neq 0\}$, are denoted by $N_{0}^k = \lvert \mathcal{I}_{0}^k \rvert$ and $N_1^k = \lvert \mathcal{I}_{1}^k \rvert$.

\subsubsection{Magnitude measures} 
For a given qubit $k$ in a $n$-qubit system, we define the \textbf{Magnitude Score}  as the difference between the total probability weight between the sectors $S_0^k$ and $S_1^k$ at a index distance $p_k$,  such that
\begin{align}
\mathrm{MS}_k = \frac{1}{2} \big( 1 + \sum\limits_{j \in S_0^k }  \;  \lvert \alpha_{j+p_k} \rvert^2 - \; \lvert \alpha_{j} \rvert^2 \big).
\end{align}
The normalization of the score is directly enforced by the sum of probabilities resulting in 1. To ensure a consistent comparison between scores, we apply a shift for it to cover the interval [0,1].
This score measures the population balance between the qubit sectors $S_0$ and $S_1$, expected to converge to $\mathrm{MS}_k \sim 0.5$ for uniformly distributed magnitudes. In contrast, if all the probability mass is either concentrated in the state $\ket{0}^{\otimes n}$ or $\ket{1}^{\otimes n}$, PS will return the values 0 or 1.

To get a more differentiated assessment, we expand the one-to-one correlation measure described above to a more general all-to-all correlations of magnitudes. Therefore we sum the pairwise product of magnitudes over both sectors S$_0^k$ and S$_1^k$, such that the \textbf{Magnitude Correlation Score} can be written as
\begin{align}
\mathrm{MCS}_k = \frac{1}{2^{n-2}} \sum_{i \in S_0^k}\sum_{j \in S_1^k} \lvert \alpha_i \rvert \lvert \alpha_j\rvert.
\end{align}

This allows us to evaluate the distribution of significant amplitude mass carried across sectors, and provides a measure of global amplitude overlap between qubit sectors.
Since the amplitude of Haar-random states follow the Porter-Thomas distribution, we expect $\mathrm{MCS}$ to approach $\pi/4 \sim 0.78$ for random states \cite{mele2024}, whereas for states that contain only mass in one of the two sectors, we expect MCS to return 0, and for uniform-amplitude states we expect it to return 1.

To measure the global localization determined by the complete probability weight distribution, we rely on the effective dimensionality. This is a fourth order-participation measure, corresponding to the amount of basis states effectively participating in a given system state. Therefore, we define the \textbf{Magnitude Dimensionality Score} for the complete $n$-qubit system as 
\begin{align}
\mathrm{MDS} = \frac{1}{D} \frac{1}{ \sum_{j = 0} ^{2^n-1}\lvert \alpha_j \rvert^4}.
\end{align}
normalized in terms of the system's dimensionality $D= 2^n$.
This score provides a direct measure of the amplitude delocalization within the system.
For a Haar-random state we expect $\mathrm{MDS}$ to converge towards $1/2$, meaning that, on average, half of all possible states participate significantly in a given state. We note here that the dimensionality score would return 1 for uniform flat-amplitude states, and $1/D$ for states with only a single basis state contribution.  

In summary, the described magnitude scores constitute three complementary measures corresponding to the first, second, and fourth moment (exponent) of the magnitude, capturing different aspects of a state's amplitude distribution. 

\subsubsection{Phase measures:} Analogously to the magnitude, we consider the phase distribution through a treatment of the pairwise angular phase comparison, relying on the relevant contributions of the $k$-th qubit across sectors $S_0^k$ and $S_1^k$ at the index distance $p_k$ respectively. We can  formulate the \textbf{Phase Score} as
\begin{align}
\mathrm{PS}_k = \frac{1}{N_{\mathrm {p}}^k}\sum\limits_{j \in \mathcal{I}_{\mathrm{p}}^k}   \; \lvert \varphi_j - \varphi _{j + p_k} \rvert
\end{align}
with the phase at the $i$-th index $\varphi_j = \mathrm{arg}( \alpha_j)$, and the number of contributing basis states $N_k$. 
In practice, we compute the phase difference modulo  $2\pi$ such that this score measures a circular distance. With this expression, we measure the one-to-one phase distance between paired qubit states differing only in qubit $k$.
For a perfectly random state we expect this score to be balanced and peaked at $0.5$, for states where every paired amplitude has a relative phase of 0, we expect it to return 0, and for a relative phase of $\pi$, we expect it to return 1.

In order to generalize phase difference between sectors to an all-to-all correlation measure, we consider the angular difference between all states across the sectors S$_0^k$ and S$_1^k$, and define  the \textbf{Phase Correlation Score} as
\begin{align}
\mathrm{PCS}_k = 1- \Biggl| \frac{1}{N_0^k N_1^k}\sum\limits_{i \in \mathcal{I}_{0}^k}  \sum\limits_{j \in \mathcal{I}_1^k}  \; \cos{( \varphi_i - \varphi _{j} )} \Biggr|.
\end{align}
This score provides a measure for cross-sector phase decorrelation.
We solely consider the phases of the elements that have a non-zero amplitude mass, to avoid contributions from empty entries. 
For a Haar-random state we expect $\mathrm{PCS}$ to converge towards 1, indicating perfect phase independence, whereas for a state with perfectly aligned phases we expect it to return 0.

To evaluate the global spread of the phase and verify its uniformity, we consider the  circular \textbf{Phase Variance Score} defined as
\begin{align}
\mathrm{PVS} = 1 - \Biggl| \frac{1}{N^k}\sum_{j \in \mathcal{I}^k} \mathrm{e}^{\mathrm{i} \varphi_j}\Biggr|,
\end{align}
with in the phase term $ \mathrm{e}^{\mathrm{i} \varphi_j} = \alpha_j/\lvert \alpha_j \rvert$.
This score measures the global phase alignment.
For Haar-random states we expect the PVS to approach 1 up to a factor $1/\sqrt{D}$, and for phase-coherent states we expect it to tend towards 0. 

In summary, the three phase scores we propose capture the phase alignment at different granularities, between cross-sector pairs, general cross-sector states or considering the states globally. 
Overall, the ensemble of scores that we propose characterizes state expressibility, including structural aspects, such as amplitude correlations and delocalization as well as phase decorrelation and variance.

\subsection{Brick-Circuit (BC)  construction for random test input state generation}
\begin{figure}
\begin{minipage}[c]{0.50\linewidth}
  \centering
  \includegraphics[width=.80\linewidth]{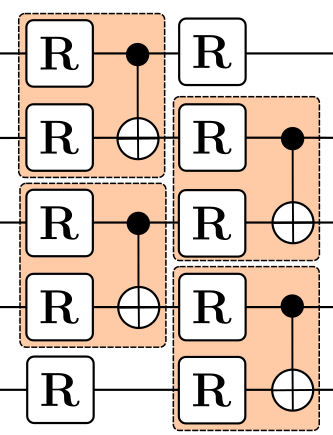}
\end{minipage} \hfill
\begin{minipage}[c]{0.45\linewidth}
\captionof{figure}{Circuit diagram of BC construction, in terms of the single-qubit random rotation $R = R_Z(\phi) R_X(\theta) R_Z(\omega)$ that can be decomposed into its Euler form, where $\phi, \omega$ are drawn randomly from the interval $[ 0, 2 \pi ]$  and $\cos\theta $ from $ [-1,1]$, and two-qubit CNOT gates, applied on alternating layers.}
 \label{fig:pattern}
\end{minipage}
\end{figure}

The diversity scores introduced above, together with the entanglement measures presented in sec.~\ref{sec:es} quantify quantum properties such as magnitude, phase, and entanglement, and together form a comprehensive set to characterize the overall diversity present in quantum test-input states. 
Furthermore, it has been shown that the expressibility (ability to explore the entire admissible state space, cf.~\ref{sec:express}) of a generated test state is directly related to its proximity to Haar randomness~\cite{sim2019,nakaji2021expressibility, azado2025expressibility}. Accordingly, we expect test circuit generators whose output states closely approximate Haar-random states to be particularly effective for expressive test input state generation, guiding our choice of the appropriate test circuit generator for QST. 

Since generating true Haar-random matrices using quantum circuits is computationally expensive, efficient generation of approximate random unitaries has been investigated in the literature. Ideas for using a two-step construction with parallel gates~\cite{emerson2003, dankert2009exact} and only hardware-compatible nearest-neighbor-gates~\cite{harrow2009, brandao2016}, as well as applying short circuit templates on alternating layers~\cite{brandao2016,cerezo2021cost, liu2022, schuster2025random} have been put forward.

In our work, we propose combining parallel local random circuits on alternating layers with unitary Haar-random single-qubit rotations into the  Brick-Circuit (BC) construction shown in Fig.~\ref{fig:pattern}, aiming to implement an efficient circuit design to generate approximate Haar-random states.
Indeed, this particular circuit construction follows the general designs that have been studied within the fields of {random matrices} and shown to efficiently approximate unitary 2-designs~\cite{emerson2003, dankert2009exact, harrow2009, brandao2016,nakaji2021expressibility,cerezo2021cost, haferkamp2022random,harrow2023approximate}, even at very shallow circuit depths~\cite{schuster2025random}. 
Our BC construction implements alternating layers of  single-qubit Haar rotations 
and entangling gates, like Controlled-Not (CNOT) or Controlled-Z (CZ), such that the circuit diagram takes the shape of a brick wall as shown in Fig.~\ref{fig:pattern}.

Note that our BC construction aligns with an established experimental construction scheme based on applying random rotation to each qubit individually and entangling gates implementing controlled gates between adjacent qubits. The choice of employed gates is compatible with the nearest-neighbour connectivity often found in hardware platforms, which enables efficient implementation of controlled gates without additional SWAP operations and ensures it is well suited for hardware characterization. Moreover, it is theoretically estimated to approximate the ideal Haar-random state with a depth that scales at most polynomial in the gate number~\cite{dankert2009exact,haferkamp2022random, harrow2023approximate, schuster2025random}.

Owing to its expected advantageous scaling of quantum property generation with a low cost of resources, we identify the BC construction as a promising candidate for test circuit generation. 
 \section{Evaluation}
\label{sec:eval}
To assess the validity of our set of diversity scores for characterizing test input states, and the effectiveness of our BC construction for generating the latter,
we formulate the following research questions (RQs):
\begin{itemize}
\item \textbf{RQ1}: To what extent do the proposed diversity scores capture the relevant state properties needed to quantify expressibility and entanglement strength of test input states?\item \textbf{RQ2}: How do various circuit generators fare in terms of circuit depth to approximate a Haar-random unitary? 
\item \textbf{RQ3}: How do different circuit generators fare in terms of the diversity of the generated test input states?
\end{itemize} 

In \textbf{RQ1} we aim to experimentally verify the capability of the theoretically established diversity scores to capture the structures in which the magnitude, phase and entanglement are present in diverse, randomized and characteristic quantum states. 

\textbf{RQ2} aims to gauge the capability of our BC generator, in comparison with existing test circuit generators, to approximate Haar-random unitaries, since  proximity to a Haar-random unitary is a commonly used criterion for evaluating randomness, uniformity and expressibility. 

Finally, \textbf{RQ3}  assesses the diversity of the test input states in terms of our established scores, where the proposed set of diversity scores enables a fine-grained, per-score analysis compared to the proximity analysis to Haar-random unitaries conducted for RQ2. The goal is to get a detailed assessment of the quantitative presence of quantum properties and hence the diversity in the test input states produced by different circuit generators.

\subsection{RQ1: Diversity Scores for evaluating state properties}
\subsubsection*{Methodology}
To answer RQ1, we analyzed the discrimination capabilities of the proposed scores with respect to the presence or lack of specific properties and structures in quantum states.
To this end, we compared the distribution of score values between randomized reference states and characteristic entangled states.  
Specifically, we evaluated the scores on four characteristic types of randomized $n$-qubit states. Therefore, we generated 5000 instances of 6-qubit states for each type, implementing Haar-random (HR), maximally-entangled bipartite (ME), uniform flat-amplitude (UA) and product (P) states, as defined in Eqs.~\ref{eq:haar}-\ref{eq:p}.
For the ME state, we chose an equal partition that divides the system evenly into two subsystems $A$ and $B$.
Additionally, we chose two well-defined entangled states, namely the GHZ and W states, as defined in Eqs.~\ref{eq:ghz}-\ref{eq:w} to ensure a complete exploration of the score's output range.  
\begin{figure}
\begin{subfigure}{.95\columnwidth}
  \centering
  \includegraphics[width=.95\linewidth]{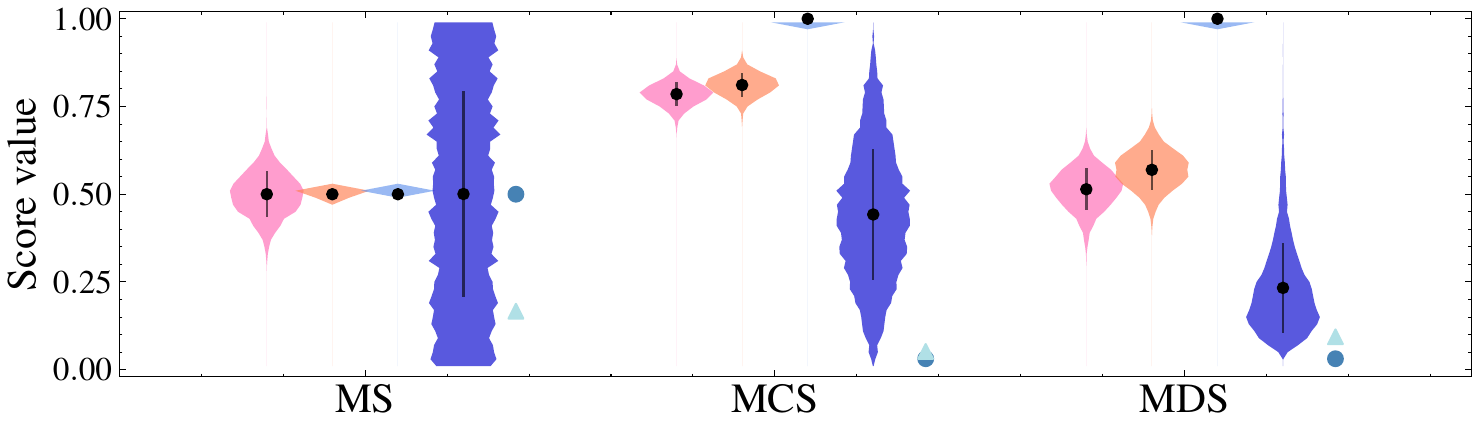}
\end{subfigure}
\begin{subfigure}{.95\columnwidth}
  \centering
  \includegraphics[width=.95\linewidth]{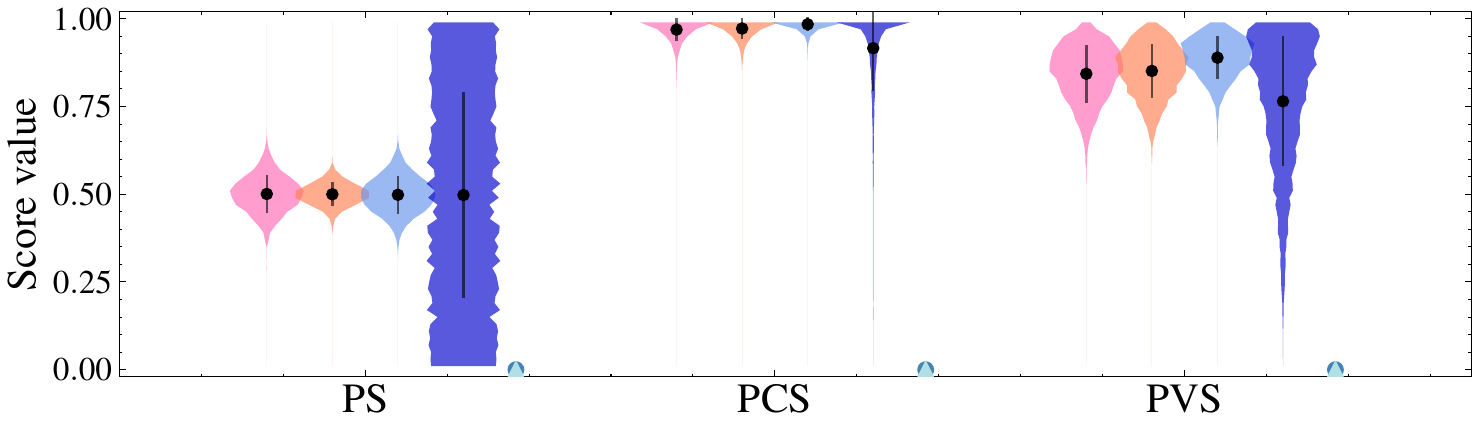}
\end{subfigure}
\begin{subfigure}{.95\columnwidth}
  \centering
  \includegraphics[width=.95\linewidth]{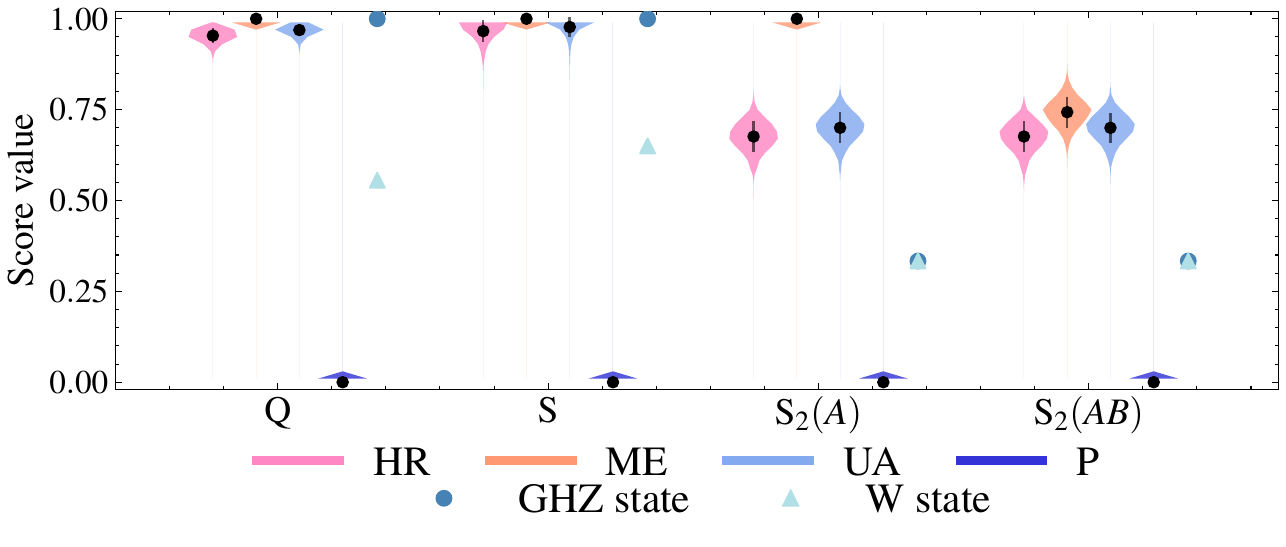}
\end{subfigure}
\caption{Scores addressing the magnitude (top), phase (middle) and entanglement (bottom) evaluated on 5,000 instances of Haar-random (HR), maximally-entangled (ME), uniform-amplitude (UA), product (P), GHZ and W states generated for 6 qubits. For the entanglement score we distinguish between R\'enyi-2 entropies that are evaluated on a block within a first subsystem $A$, and a block across the partition $A \lvert B$.}
\label{fig:state_scores}
\end{figure}

For the entanglement scores, 
we consider the Meyer-Wallach measure Q (cf. Eq.~\ref{eq:qmw}), the von Neumann entropy (S) (cf. Eq.~\ref{eq:entropy}), and the R\'enyi-2 entropy (S$_2$) as defined in Eq.~\ref{eq:r2e},
where the latter is evaluated on two different subsystems. 
We chose a certain partition of the global state on which we evaluate S$_2$ by clustering the considered qubits into blocks. By S$_2(A)$ we denote the entropy that is evaluated on a block fully contained within subsystem $A$ of a fixed bipartition  $A\lvert B$, whereas S$_2(AB)$ is evaluated on a block that includes the partition between the subsystems $A$ and $B$.
To ensure comparison, during the evaluation, S$_2$ is normalized by the size of the block under consideration, and thus returns the R\'enyi‑2 entropy per qubit in the block.

\subsubsection*{Results}
The resulting distributions of the evaluated scores are shown in Fig.~\ref{fig:state_scores}, grouped by the three properties that they are addressing, namely magnitude (top), phase (middle), and entanglement (bottom). 

As shown in Fig.~\ref{fig:state_scores} (top), for \textbf{MS}, we observe that the HR, ME,  P and GHZ states
are centered around 0.5, indicating an equilibrated balance between qubit sectors, as expected from states that are symmetric in each qubit's $\ket{0}$ and $\ket{1}$ state. By construction, the W state has a larger amplitude mass in the $\ket{0}$ states, which is confirmed by its lower MS value  (MS = 0.17). The widest distribution is achieved by the product state, due to its random instantiation of each individual qubit. In this case, each qubit is sampled independently and uniformly, implementing a Haar measure on the single qubit pure states, leading to a uniformly distributed probability where no constraints enforce concentration, captured by MS, defined as a sector-wise score. 
Considering \textbf{MCS} and \textbf{MDS}, the UA states achieve the maximal score, since by design they include contributions from all possible basis states. 
In contrast, the W and GHZ state with very few basis state contributions result in a minimal value of the score. 
In terms of amplitude distribution, the ME states show a high similarity to the HR states. 
The retrieved average expected value of the MCS for the HR states results in 0.78 and its MDS value agrees with the expected 0.5. 
The P states show a moderate MCS and MDS, expected by the factorization of the probability mass.

Fig.~\ref{fig:state_scores} (middle) shows that the distributions obtained for the \textbf{PS} closely resemble those observed for the \textbf{MS}, with all randomized states lying around the mean value indicating balanced phases, and the product states spanning the entire range from $0$ to $1$. For the \textbf{PCS} and \textbf{PVS} all of the randomized states generate strongly decorrelated phases and a large phase spread. In contrast, the W and GHZ states correctly result in low phase score values since their phases are completely aligned by design. The pronounced tail for the P states reflects their more limited ability to generate strong phase-scrambling.

Fig.~\ref{fig:state_scores} (bottom) confirms that all entanglement scores correctly assign zero entanglement to the product states, as required for any valid entanglement measures. 
For both the \textbf{Q} and \textbf{S} scores, HR, ME, UA and the GHZ states exhibit distributions that are strongly peaked near their maximal values, confirming that on average each qubit is strongly entangled with the remainder of the system. 
The W state, by contrast, shows only moderate entanglement according to Q and S ($\sim 0.6$), consistent with its delocalization of a single excitation over many qubits and the resulting weak single‑qubit entanglement.

A qualitatively different behavior emerges when considering the R\'enyi‑2 entropy \textbf{S$_2$}, which we expect to probe the scaling of entanglement with subsystem size.
For the ME states, we observe a pronounced reduction in entanglement when evaluating blocks that cross the state construction bipartition $A\lvert B$, as captured by S$_2(AB)$. Indeed, while S$_2(A)$ reaches its maximal value of 1.0,  S$_2(AB)$ returns a mean value of 0.74. This behavior is expected, since volume‑law entanglement is realized only within block $A$ by design.
For both  GHZ and W states, S$_2$ yields a value of $\frac{1}{3}$ independent of the chosen block, consistent with the presence of one global bit of entanglement shared across the 3-qubit subsystems. 
In contrast, HR and UA states achieve similarly high values for Q and S, but moderately reduced values for S$_2(A)$ and  S$_2(AB)$, consistent with the range of entanglement which is shorter than that of the ME states. 
The independence of the entanglement strength on the block choice  for HR and UA states confirms an isotropic entanglement distribution, while the S$_2$ values, significantly exceeding those of GHZ and W state, indicate a favorable extensive entanglement scaling in the randomized ensembles.

These results confirm that the diversity scores we introduced, addressing magnitude, phase, and entanglement, are capable of differentiating between the structural characteristics of the states considered. 
Indeed, the selected entanglement measures recover the expected entanglement signatures of well‑understood reference states and reliably distinguish structured entanglement from that arising in randomized ensembles.

\vspace{0.15cm}
\noindent
\setlength{\fboxsep}{5pt}\setlength{\fboxrule}{1pt}\fcolorbox{blue!95}{blue!10}{\parbox{0.945\columnwidth}{\textbf{Answer to RQ1.} The set of the proposed diversity scores capture well the different properties present in the quantum reference states we considered, resulting in good agreement with the underlying theory.}
    }
\vspace{0.15cm}

\subsection{RQ2: Randomness Analysis}
\label{sec:eval2}

\subsubsection*{Methodology}
\textbf{RQ2} addresses the difficulty of generating a true Haar-random unitary using a finite set of discrete gates without encountering exponentially increasing circuit depth with the system size. To this end, we implement the BC construction and compare the amount of gates required to generate circuit matrices that are Haar-random beyond statistical distinguishability with three existing test circuit generators.

We employ the statistical test of maximum mean discrepancy (MMD) on generated circuit matrices to evaluate their proximity to a Haar-random matrix.
The MMD hypothesis test is used to compare distances between the embeddings of two distributions given by their unitary matrices~\cite{borgwardt2006integrating}. We attribute a p-value to conclude whether the difference between the realization of the distributions is statistically insignificant and adopt a threshold value of $p > 0.01 $.

In practice, ideal Haar unitary matrices are obtained via the standard QR decomposition method~\cite{mezzadri2007generate}. We generate a set of these ideal random matrices,
as well as circuit matrices that are produced by our BC generator.
In addition, our experiments include three existing test circuit generators, i.e., Random (RC), UCNOT and iQFT,  proposed by QuraTest~\cite{ye2023qura}, a state-of-the-art approach in quantum software testing.

We employed the PennyLane framework~\cite{bergholm2018pennylane} to generate quantum circuits and extract their circuit matrix representation.
To vary the size of the considered system we generate circuit matrices for 3, 5 and 7 qubits.

\subsubsection*{Results}
\begin{figure}[tb]
  \centering
  \includegraphics[width=.95\linewidth]{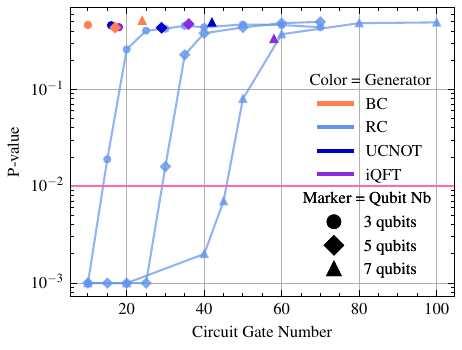}
  \caption{P-value comparison of Random Circuit (RC), Brick Construction (BC), UCNOT and iQFT circuit generators for 3, 5 and 7 qubits and varying circuit depth.}
  \label{fig:p_value_qubit_nb}
\end{figure}

Fig.~\ref{fig:p_value_qubit_nb} shows how circuit depth influences the randomness of circuit matrices generated for three different system sizes, namely including 3, 5 and 7 qubits, by the different circuit constructions. 
Considering the case for 5 qubits, circuit unitaries generated using the \textcolor{BC}{BC} construction
already satisfy our randomness criterion at a single-layer depth, corresponding to circuits with 17 gates. 
In contrast, the \textcolor{RC}{RC} generator requires approximately 35 gates to achieve a comparable approximation to a Haar‑random unitary.  The observed flattening of the curve is consistent with the the expected saturation of expressibility~\cite{sim2019}.

The \textcolor{UCNOT}{UCNOT} generator also performs well in producing random states; we illustrate the results obtained for one single layer, since increasing the number of layers 
does not lead to a noticeable improvement. Even its shallowest configuration, which reaches the randomness threshold, requires 29 gates.

Finally, the \textcolor{iQFT}{iQFT} generator has a fixed circuit depth determined by its algorithmic structure, which corresponds to 36 gates for a 5‑qubit system. As expected, the resulting unitaries exhibit lower randomness, since the iQFT involves fewer random parameters and is designed to produce a more structured output.

 As expected, with increasing qubit number a higher gate count to satisfy the randomness criterion is required. We observe that the scaling advantage of the BC construction scheme is reinforced with larger system sizes compared to the other circuit construction methods, as can be seen from the more slowly increasing amount of required gates with growing system size listed in Table~\ref{tab:gate_requirements}. 

\begin{table}[t]
\caption{Required gate numbers to reach the randomness criterion for different test circuit generators and system sizes.}
\label{tab:gate_requirements}
\centering

\begin{tabular}{c|cccc}
\hline
\textbf{Number of Qubits} & \textbf{BC} & \textbf{RC} & \textbf{UCNOT} & \textbf{iQFT} \\
\hline
3  & 10 &  15 & 16  & 18 \\
5  & 17 & 30 & 29  & 36 \\
7  & 24 & 50 & 42 & 58 \\
\hline
\end{tabular}
\end{table}

\vspace{0.15cm}
\noindent
\setlength{\fboxsep}{5pt}\setlength{\fboxrule}{1pt}\fcolorbox{blue!95}{blue!10}{\parbox{0.945\columnwidth}{\textbf{Answer to RQ2.} The BC generator fulfills the randomness criterion with the least amount of gates, followed by the UCNOT, iQFT generators, and finally the RC generator.}
    }
\vspace{0.15cm}

\subsection{RQ3: Diversity Scores for expressibility and entangling capabilities}
\subsubsection*{Methodology}
To address \textbf{RQ3}, we evaluate test input states generated by the BC construction and three existing test‑case generators using our established scores. 
Therefore, we apply the generated circuit matrices onto the default initial state in order to generate test input states, and assess the diversity of the resulting states in terms of expressibility and entangling capability.

Therefore, we generated 2000 instances of circuit matrices for each of the four generators (BC, RC, UCNOT, iQFT) for a 5-qubit system. We explored the capability of the BC and RC generators to allow for a variable gate number to study the diversity of the generated test input states with increasing circuit depth.
Additionally, we computed 2000 instances of Haar-random states, representing a uniform state distribution on the whole multi-qubit state space and thus acting as a baseline for expressibility.
We evaluated each of these with the proposed scores targeting magnitude, phase, and entanglement distributions. 
We used the Wasserstein metric to quantify the statistical distance between the distributions obtained from evaluating the diversity scores on test input states constructed by the different generators and on Haar-random states. 

\subsubsection*{Results}
\begin{figure}
\begin{subfigure}{.98\columnwidth}
  \centering
  \includegraphics[width=.95\linewidth]{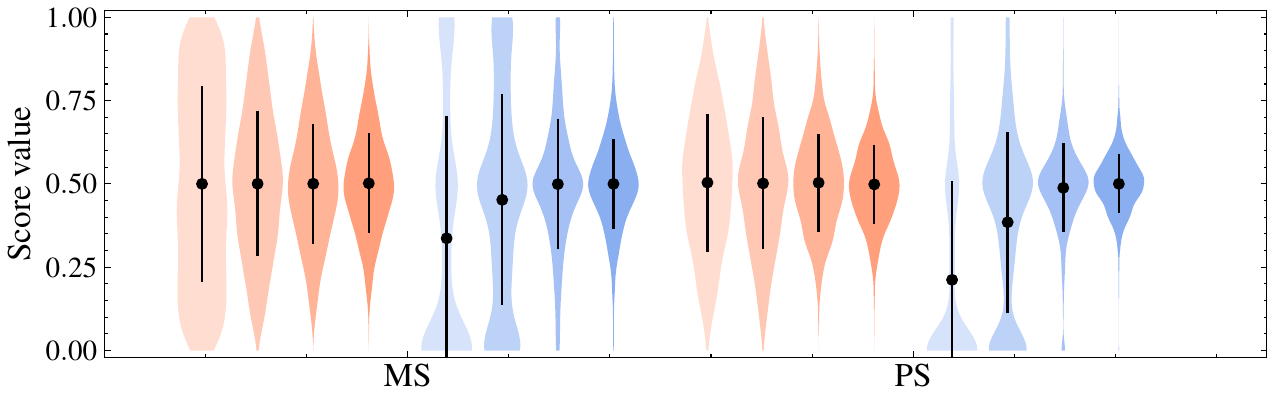}
\end{subfigure}
\begin{subfigure}{.98\columnwidth}
  \centering
  \includegraphics[width=.95\linewidth]{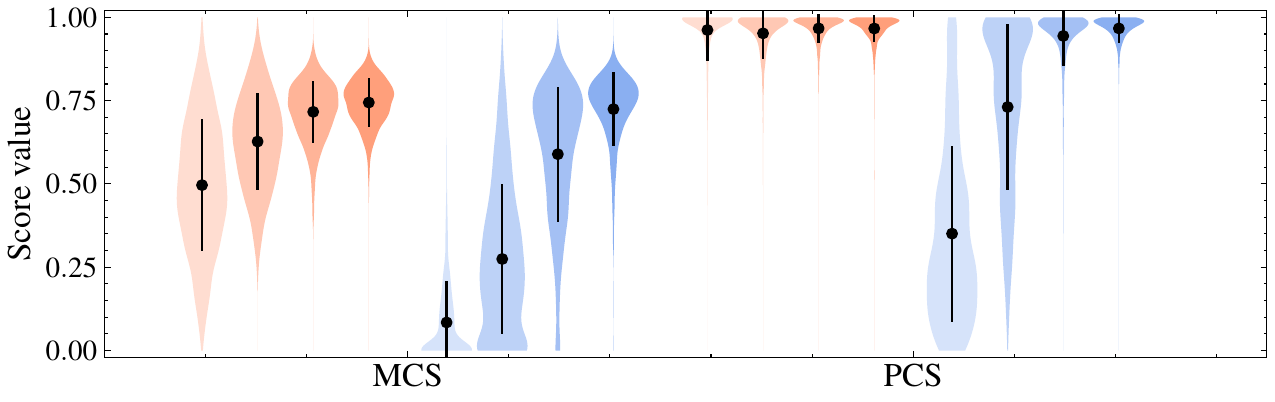}
\end{subfigure}
\begin{subfigure}{.98\columnwidth}
  \centering
  \includegraphics[width=.95\linewidth]{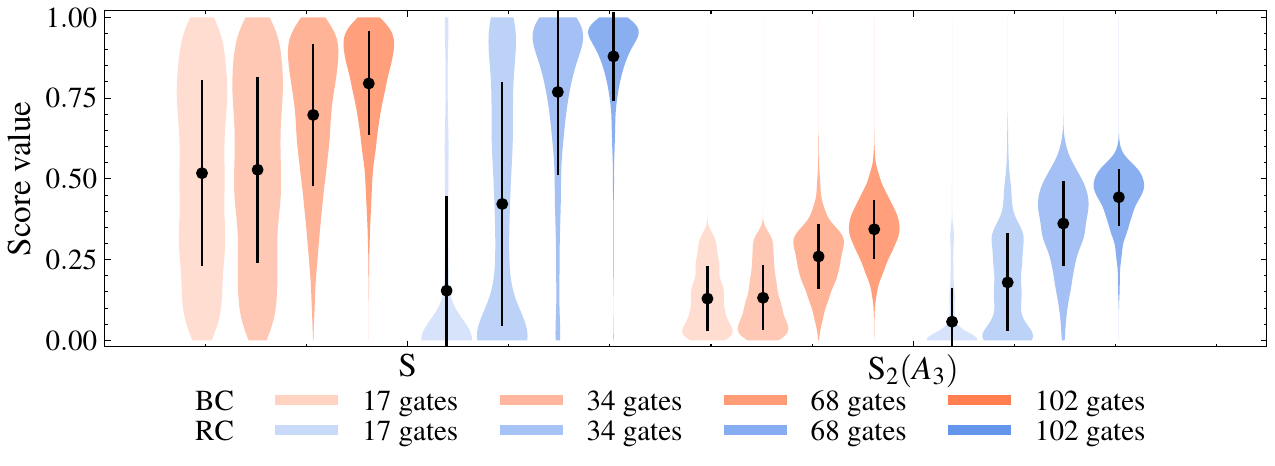}
\end{subfigure}
\caption{Diversity Scores for increasing gate numbers. Magnitude (MS) and phase (PS) scores (top), magnitude (MCS) and phase (PCS) correlation scores (middle) and entanglement entropy S and R\'enyi-2 entropy evaluated on a 3-qubit block S$_2(A_3)$ (bottom) for 2000 instances of a 5-qubit system with increasing gate numbers, corresponding to 1, 2, 4 and 6 layers of the BC construction.}
\label{fig:violin_gate_nb}
\end{figure}

\begin{figure}[tb]
  \centering
  \includegraphics[width=.95\linewidth]{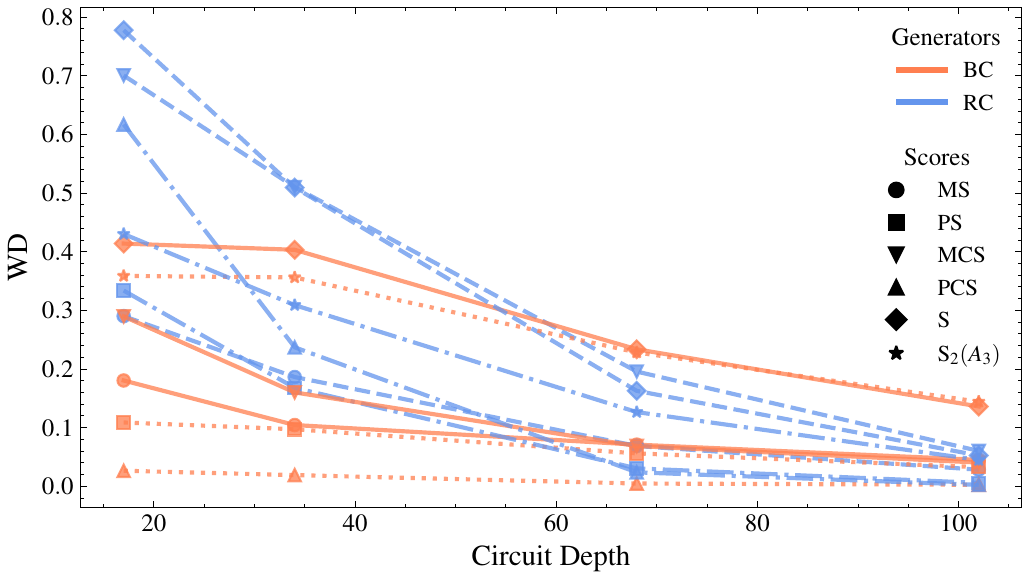}
  \caption{Wasserstein Distance (WD) of the diversity scores evaluated on the test input states generated by BC and RC constructions with varying circuit depth with respect to Haar-random reference scores.}
  \label{fig:stat_dist}
\end{figure}

First, we analyze the varying distribution of magnitude and phase in the generated test circuits with increasing circuit depth. In Fig.~\ref{fig:violin_gate_nb}, we show the magnitude and phase scores MS and PS (top), the magnitude and phase correlation scores MCS and PCS (middle), and entanglement scores S and S$_2(A_3)$ (bottom) for the RC and BC construction  with increasing gate numbers. 

We observe a clear variation in the shape of the distribution with increasing circuit depth for the considered generators and scores. For \textbf{MS} and \textbf{PS} in the case of the BC construction the mean value of the score remains constant around the balanced value 0.5, however, the distribution  becomes more narrowly concentrated, in agreement with the expectations for Haar-random states. 
For the RC generator, the change with increasing allowed gate number is more drastic. In this case also the mean value, so the average generated magnitude (\textbf{MS} from 0.34 to 0.5) and phase (\textbf{PS} from 0.21 to 0.5) weight increase significantly, indicating broadened phase and magnitude diversity.  A very similar picture is obtained for the correlation scores, where for \textbf{MCS} both generators produce increasingly correlated values with the gate number, whereas for \textbf{PCS}, the BC produces maximally phase decorrelated state already with shallow circuits, whereas RC requires a significant number of gates ($\sim100)$ to fulfill this condition. 
For the entanglement scores, \textbf{S} and \textbf{S}$_2(A_3)$, both generators approach the maximal entanglement score, with increasingly more narrow distributions as circuit depth grows. We even observe that for deep circuits, the states produced by the unstructured RC generator outperform the values obtained for the structured BC construction. Furthermore, both generators achieve a higher mean value for S, 0.8 and 0.88 for BC and RC respectively, than for S$_2(A_3)$, which is limited to 0.34 and 0.44 for BC and RC. This illustrates again (cf. previous section) that it is significantly more difficult to generate genuinely multipartite, delocalized entanglement compared to single-qubit entanglement, especially for structured generators.

In comparison to the results from RQ2, we note that our diversity scores provide a more detailed tool to assess the expressibility of the generated test circuits, rather than solely considering their distinguishability in terms of randomness. Indeed, we observe that the generated circuits are becoming more diverse (achieving higher scores, which lie closer to the Haar-random value) even for gate number above 17 for the BC construction and 30 for RC that exceeded the randomness threshold set in the previous section. 

By increasing the number of allowed gates, especially for the RC construction, the distributions of the considered scores progressively approach that obtained from the Haar-random states,  as illustrated by the statistical distance plotted in Fig.~\ref{fig:stat_dist}. Smaller distances indicate a closer match to the Haar-random ensemble and therefore higher expressibility of the corresponding circuit generator.  
Furthermore, we observe that a gate depth of 17 and below is insufficient for the RC construction to reproduce the state properties that we expect for random circuits.

\begin{figure}
\begin{subfigure}{.98\columnwidth}
  \centering
\includegraphics[width=.95\linewidth]{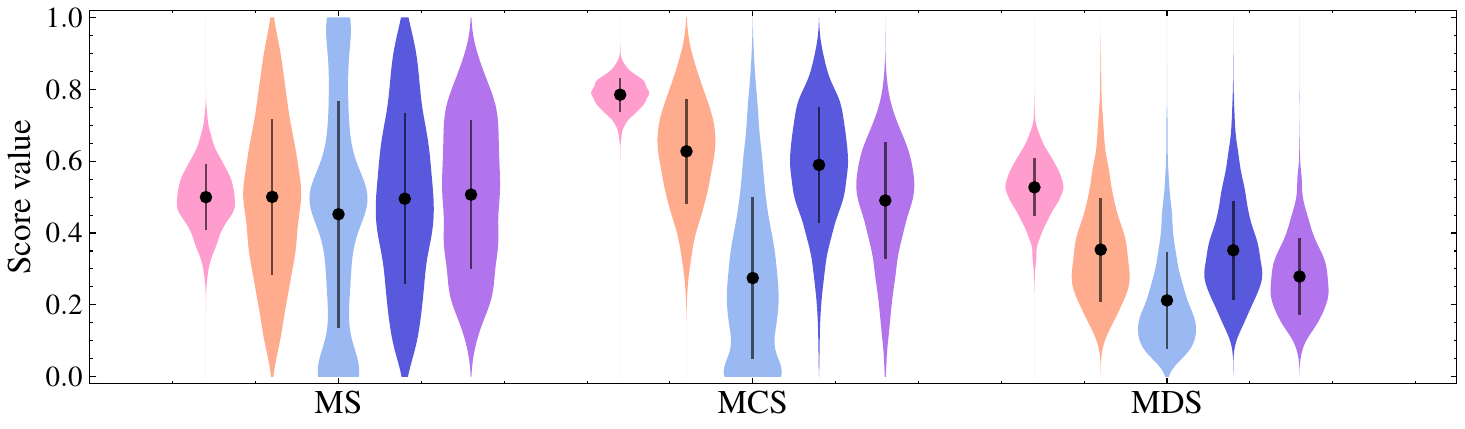}

\end{subfigure}
\begin{subfigure}{.98\columnwidth}
  \centering
\includegraphics[width=.95\linewidth]{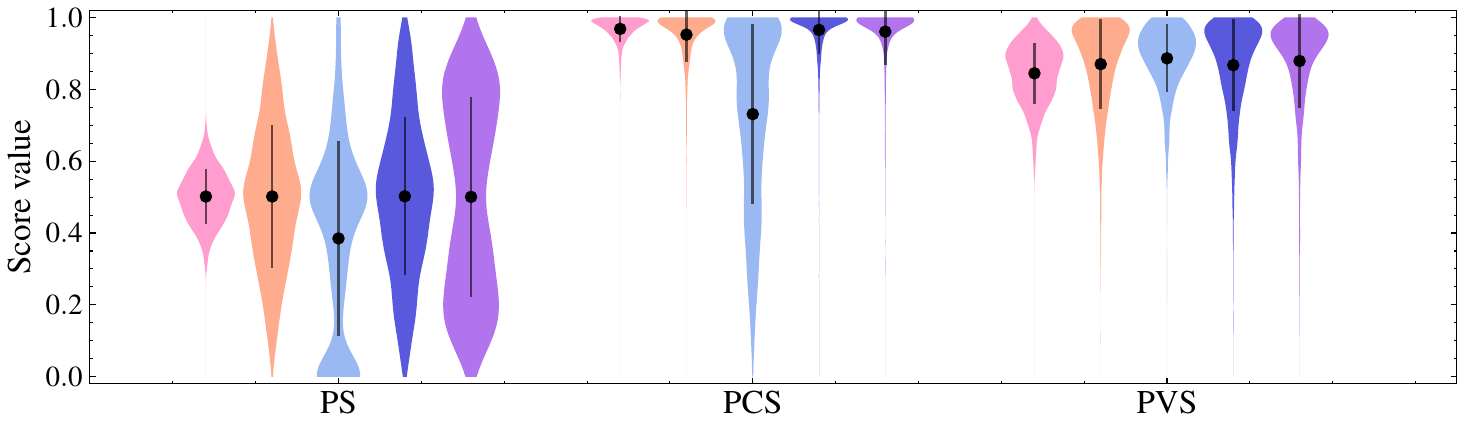}
\end{subfigure}
\begin{subfigure}{.98\columnwidth}
  \centering
\includegraphics[width=.95\linewidth]{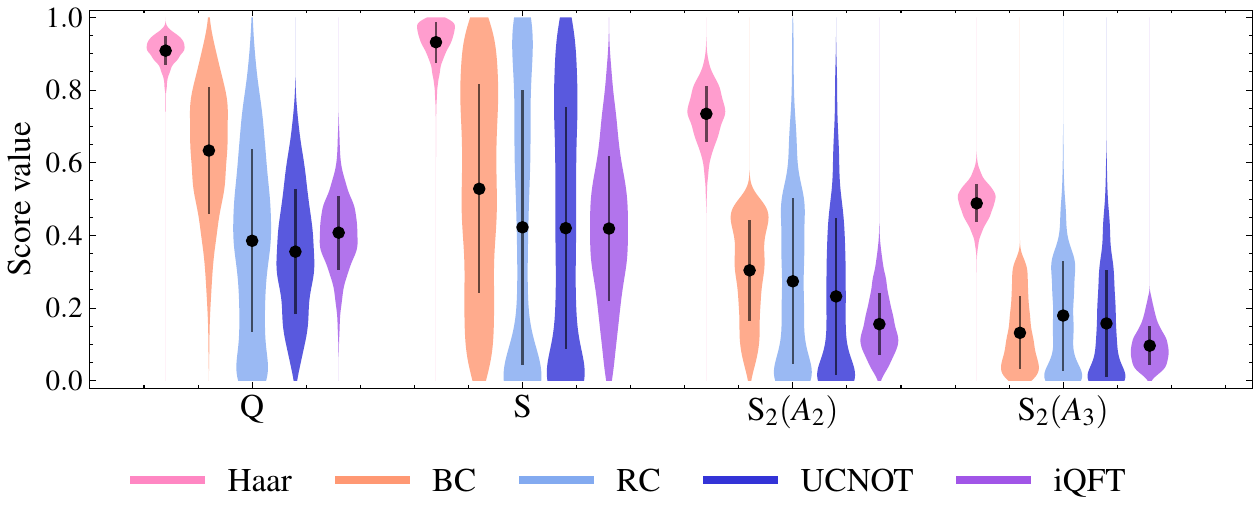}
\end{subfigure}
\caption{Diversity scores for magnitude (top), phase (middle) and entanglement (bottom) evaluated on test input states produced by BC (2 layers), RC (34 gates), UCNOT and iQFT generators, compared to Haar-random states for 5 qubits. 
}
\label{fig:violin_unitary_scores}
\end{figure}

The complete overview of the score distributions  obtained for states originating from different generators is shown in Fig.~\ref{fig:violin_unitary_scores}. 
To allow for a fair comparison between generators,  
the RC and BC constructions are evaluated on circuit depths that are similar to the fixed gate generators UCNOT and iQFT (2 layers for the BC, corresponding to 34 gates for the RC).

For \textbf{MS}, all test states are symmetric around \textbf{MS} $\sim0.5$, and thus show balanced two-sector states. For the \textbf{MCS} and  \textbf{MDS}, the BC achieves the highest scores (closest to Haar-random prediction), followed by the UCNOT, iQFT and RC construction.
As for the \textbf{PS}, all generators except the RC are balanced around 0.5 and are capable of generating states with completely decorrelated phases \textbf{PCS} $\sim 1$, whereas all generators achieve a uniform phase spread captured by \textbf{PVS} close to its maximal value.
For the entropy \textbf{S} the BC construction achieves the highest mean score, closest to the Haar-random distribution; all other generators result in very similar mean values, whereas, with the more narrow distribution, the iQFT generator is slightly advantageous in its reliability to generate entanglement compared to RC and UCNOT. 
In the case of the Meyer-Wallach measure \textbf{Q}, for which the mean value  $\overline{Q}$ can be directly related to the entangling capability~\cite{sim2019}, the score results in the average value $\overline{Q} = 0.63$ for the BC construction, 0.41 for iQFT, 0.39 for RC and 0.36 for UCNOT, in comparison to the Haar-random state which achieves  $\overline{Q} =0.9$.
In fact, the computed WDs of the Q score with respect to Haar-random states result in 0.27 for the two layer BC construction, 0.52 for the RC generator, 0.55 for UCNOT and 0.50 for iQFT. For the R\'enyi-2 entropy \textbf{S$_2$}, for the 2-qubit block, the BC scores the strongest entanglement, followed by RC, UCNOT and iQFT generators. For the larger 3-qubit block, the RC slightly outperforms the other circuit generators, followed by the very similar distribution generated by UCNOT, BC and iQFT. 
The larger values obtained for S in comparison to S$_2$, and especially their reduction with increasing block size, indicate that although single-qubit entanglement is strong, the entanglement  is limited to a short range and localized. We also note that the qualitative features of the distributions that are captured by $S$ are also present in the computationally advantageous $Q$ measure, which is preferable for practical applications.

\vspace{0.15cm}
\noindent
\setlength{\fboxsep}{5pt}\setlength{\fboxrule}{1pt}\fcolorbox{blue!95}{blue!10}{\parbox{0.945\columnwidth}{\textbf{Answer to RQ3.} The BC construction achieves the highest diversity scores approaching the baseline value we obtain for Haar-random states in all three aspects, namely magnitude, phase and entanglement, followed by the UCNOT, iQFT and RC generators.}
    }
\vspace{0.15cm}

\subsection{Threats to Validity}
\label{sec:val}
To reduce the threats of random variation and insufficient statistical power, we employ at least 2000 samples of each generator, and used a common statistical test, such as the maximum mean discrepancy, and the Wasserstein distance between the score distribution of generated test input states and the Haar-random reference to analyze the results. However, due to the inherent variability of the randomness involved, some results concerning the state scores may still be affected by statistical outliers and finite-size effects.

Our diversity scores address quantities that are not direct physical observables, and as such they are not uniquely defined.  To mitigate this threat, we selected three different measures to capture different aspects of each property and apply them consistently across all evaluations. \section{Related Work}
\label{sec:rel}
In the context of quantum software testing, test case generation requires quantum input states to be produced by a quantum circuit. 
Prior work, most notably QuraTest~\cite{ye2023qura} has investigated the use of quantum-inspired measures to assess the diversity of test inputs generated by different circuit constructions. 
In this setting, diversity is quantified by three scores probing the magnitude, phase and entanglement across a bipartition, effectively capturing average cross-sector correlations associated with individual qubit being in state $\ket{0}$ or $\ket{1}$. 
In contrast, our diversity scores provide a more comprehensive characterization by explicitly targeting state-space exploration, rather than limiting the analysis to a single-cut correlation measure. 
Furthermore, these scores are used to evaluate the performance in input and output state coverage of three test circuit generators, consisting of a Random gate combination, a UCNOT pattern and an IQFT circuit. 
In contrast to our BC generator, none of these existing test circuit generators is specifically optimized for circuit depth efficiency or for producing strongly entangled, highly expressive states.
To the best of our knowledge, despite its highly desirable properties for generating diverse and comprehensive test inputs, this work constitutes the first systematic investigation of the BC construction as a test circuit generator for QST.

Research on quantum random circuits spans a wide range of applications, including complex quantum many-body and back-hole physics~\cite{nahum2018operator, hayden2007black}, device benchmarking~\cite{emerson2005scalable,magesan2011scalable,cross2016random}, variational quantum algorithms~\cite{sim2019,nakata2021quantum,nakaji2021expressibility}, quantum machine learning~\cite{mcclean2018barren,liu2023analytic}, and near-term demonstrations of quantum advantage~\cite{boixo2018characterizing,liu2022,arute2019supremacy}. 
Since implementing ideal Haar-random unitaries using discrete gates is computationally expensive, practical approaches focus on efficiently approximating Haar randomness up to an error~$\epsilon$, formalized through the framework of unitary $t$-designs~\cite{harrow2009,brandao2016}.  
Accordingly, a central challenge resides in the practical construction of pseudo-random circuits that approximate Haar properties with minimal resources.
To this end, numerous efficient schemes for approximate unitary design have been investigated for their convergence to unitary $t$-designs at shallow circuit depth ~\cite{emerson2003,harrow2009,dankert2009exact,brandao2016,nakata2021quantum,cerezo2021cost,haferkamp2022random,liu2022,harrow2023approximate,schuster2025random}.
In contrast, the BC generator introduced in this work instantiates the established parallel local random-circuit paradigm with a specific aim at efficient QST, combining low gate counts with nearest-neighbor entangling gates, making it well-suited for implementation on real-world hardware~\cite{nahum2017quantum, boixo2018characterizing, arute2019supremacy}.

Random-circuit designs are especially relevant for randomized device benchmarking, where it is used to estimate process fidelities and characterize noise in quantum processors~\cite{emerson2005scalable,magesan2011scalable, cross2016random} by generating constructible families of states that approximate uniform sampling over the full Hilbert space. Related ideas also appear in variational quantum algorithms, where expressibility serves as a key metric characterizing how well a given ansatz can cover the Hilbert space~\cite{sim2019,nakaji2021expressibility, azado2025expressibility}.
Building on these prior studies, our work leverages the established link between expressibility and entangling capabilities of parametrized quantum circuits, as well as their proximity to Haar‑distributed ideal random states to assess the diversity of test input states using the proposed diversity scores. 
This plays a central role in QST, since the Hilbert space coincides with the set of all admissible input states and its extensive exploration forms the basis of any comprehensive testing scheme.

\section{Conclusions}
\label{sec:con}
In this work, we have proposed a set of diversity scores to guide test-case generation for comprehensive quantum software testing, and paired them with our BC construction method for efficient random test circuit generation. 
We have addressed the challenge to generate a diverse set with high coverage of testing inputs, by taking the general state space structure of quantum mechanical systems into account and considered the randomness of the generated test cases. 
We have compared the expressibility and entangling capabilities of differently structured test circuit generators and observed that, from the considered implementations, our BC construction reaches the highest scores, with the most shallow gate depth.  

As part of future work, one can refine the analysis of the circuit generator's required gate depth by distinguishing between single and two-qubit gates.
Furthermore, we plan to study the impact of an expressible and entangling test circuit generator on the efficacy of the fault finding process, for instance, leveraging fault injection. 
These efforts will further enable the effective generation and use of random circuits in QST. 

\section*{Data Availability Statement}

A replication package containing all the code and the experimental data is provided online~\cite{qstdata}.
We plan to make it publicly available upon acceptance.

\bibliographystyle{IEEEtran}

\end{document}